\documentclass[11pt,twoside]{article} \usepackage[margin=1in]{geometry}
\usepackage{amssymb}\usepackage[unicode,pdftex]{hyperref}\begin{document}

\newcommand\hreff[1]{\href {http://#1} {\small http://#1}}
\newcommand\trm[1]{{\bf\em #1}}\newcommand\prf{\paragraph{Proof.}}
\newtheorem{thr}{Theorem} \newtheorem{lem}{Main Lemma}
\newtheorem{prp}{Proposition}\newcommand\emm[1]{{\ensuremath{#1}}}
\newcommand\qed{\hfill{\emm\blacksquare}}

\newcommand\floor[1]{{\lfloor#1\rfloor}}\newcommand\ceil[1]{{\lceil#1\rceil}}
\newcommand\lea{\prec}\newcommand\gea{\succ}\newcommand\eqa{\asymp}
\newcommand\lel{\lesssim}\newcommand\gel{\gtrsim}
\newcommand\edf{{\,\stackrel{\mbox{\tiny def}}=\,}} \newcommand\ov{\overline}

\newcommand\N{{\emm{\mathbb N}}}\newcommand\tb{{\mathbf t}}
\newcommand\K{{\mathbf K}} \newcommand\I{{\mathbf I}}
\newcommand\M{{\mathbf M}}\renewcommand\d{{\mathbf d}}
\newcommand\Ks{{\raisebox{2pt}{$\chi$}}} \newcommand\St{{\mathbf S}}
\newcommand\ch{{\mathcal H}} \renewcommand\l{\lambda}

\author {Leonid A.~Levin\\ Boston University\thanks
 {Computer Sci.\@ dpt., 111 Cummington Mall, Boston, MA 02215.
 My homepage: \hreff{www.cs.bu.edu/fac/lnd}}}
\title{\vspace*{-3pc} Occam Bound on Lowest Complexity of Elements.\thanks
 {This research was supported in part by NSF grant CCF-1049505.}}

 \date{}\maketitle\begin{abstract}\noindent
 The combined universal probability $\M(D)$ of strings $x$ in sets $D$
is close to $\max_{x{\in}D}\M(\{x\})$: their $\sim\log$s differ by at most
$D$'s information $j{=}\I(D:\ch)$ about the halting sequence $\ch$.
 Thus if all $x$ have complexity $\K(x)\ge k$, $D$ carries $\ge i$ bits
of information on each $x$ where $i{+}j\sim k$. Note, there are no
ways (whether natural or artificial) to generate $D$ with significant
$\I(D:\ch)$. \end{abstract}

\section {Introduction.}

Many intellectual and computing tasks require guessing the hidden part of the
environment from available observations. In different fields these tasks have
various names, such as Inductive Inference, Extrapolation, Passive Learning,
etc. The relevant part of the environment can be represented as an, often huge,
string $x{\in}\{0,1\}^*$. The known observations restrict it to a set $D\ni
x$.\footnote
 {$D$ is typically enormous, and a much more concise theory can
  often represent the relevant part of what is known about $x$.
  Yet, such {\em ad hoc} approaches are secondary:
  raw observations are anyway their ultimate source.}

One popular approach to guessing, the ``Occam Razor,'' tells to focus on
the simplest members of $D$. (In words, attributed to A. Einstein, ``A
conjecture should be made as simple as it can be, but no simpler.'')
Its implementations vary: if two objects are close in simplicity, there
may be legitimate disagreements on which is slightly simpler. This
ambiguity is reflected in formalization of ``simplicity'' via the
Kolmogorov Complexity function $\K(x)$ - the length of the shortest
prefix program\footnote
 {This analysis ignores issues of finding short programs efficiently.
Limited-space versions of absolute complexity results are usually
straightforward. Time-limited versions often are not, due to
difficulties of inverting one-way functions. However the inversion
problems have time-optimal algorithms. See such discussions in \cite{uh}.}
 generating $x$: $\K$ is defined only up to an additive constant
depending of the programming language. This constant is small compared
to the usually huge whole bit-length of $x$. More mysterious is the
justification of this Occam Razor principle.

A more revealing philosophy is based on the idea of ``Prior''. It
assumes the guessing of $x\in D$ is done by restricting to $D$ an {\em a
priori} probability distribution on $\{0,1\}^*$. Again, subjective differences
are reflected in ignoring moderate factors: say in asymptotic terms,
priors different by a $\theta(1)$ factors are treated as equivalent.
 The less we know about $x$ (before observations restricting $x$ to $D$)
the more ``spread'' is the prior, {i.e.} the smaller would be the variety
of sets that can be ignored due to their negligible probability.
 This means that distributions truly prior to any knowledge, would be the
largest up to $\theta(1)$ factors. Among enumerable ({i.e.} generatable as
outputs of randomized algorithms) distributions, such largest prior does in
fact exist and is $\M(\{x\})=2^{-\K(x)}$.

These ideas developed in \cite {S} and many subsequent papers do remove
some mystery from the Occam Razor principle. Yet, they immediately yield
a reservation: the simplest objects have {\bf each} the highest
universal probability, but it may still be negligible compared to the
{\bf combined} probability of complicated objects in $D$. This suggests
that the general inference situation might be much more obscure than the
widely believed Occam Razor principle describes it.

\paragraph {The present paper} shows this could not happen, except as a purely
mathematical construction. Any such $D$ has high information $\I(D:\ch)$
about Halting Problem $\ch$ (``Turing's Password'' :-). So, they are
``exotic'': there are no ways to generate such $D$; see this informational
version of Church-Turing Thesis discussed at the end of \cite{fi}.

Consider finite sets $D$ containing only strings of high ($\gel k$)
complexity. One way to find such $D$ is to generate at random a small
number of strings $x\in\{0,1\}^k$. With a little luck, all $x$ would
have high complexity, but $D$ would contain virtually all information
about each of them.

Another (less realistic :-) method is to gain access to the halting
problem sequence $\ch$ and use it to select for $D$ all strings $x$ of
complexity $\sim k$ from among all $k$-bit strings. Then $D$ contains
little information about most of its $x$ but much information about $\ch$ !

Yet another way is to combine both methods. Let $v_h$ be the set of all
strings $vs$ with $\K(vs)\sim\|vs\|{=}\|v\|{+}h$. Then $\K(x)\sim i+h$,
$\I(D:x) \sim i$, and $\I(D:\ch)\sim h$ for most $i$-bit $v$ and
$x{\in}D=v_h$. We will see no $D$ can be better: they all contain strings
of complexity $\lel\min_{\,x\in D} \I(D:x){+}\I(D:\ch)$.

\vspace{1pc} The result is a follow-up to Theorem~2 in~\cite{VV10}.
\cite{VV04} provides in Appendix I more history of the concepts used here;
\cite{K,S,LV08} give more material on Algorithmic Information Theory.
 {{This work's central idea is due to S. Epstein, appearing in \cite{EB11}.
    He is a co-author of an earlier preprint \cite{EL} of the results
    below and a sole author of their many extensions in \cite{ep-d}.}}

\section {Conventions and Kolmogorov Complexity Tools.}

 $\|x\|{\edf}n$ for $x{\in}\{0,1\}^n$; for $a{\in}\Re^+$,
 $\|a\|{\edf} \ceil{\,|\log a|\,}$. $\St{\edf}\{0,1\}^*$.
 $p0^-{=}p1^-{\edf}p$ ; $\emptyset^-$ is undefined.
 $[A]\edf1$ if statement $A$ holds, else $[A]\edf0$.
 ${\lea}f$, ${\gea}f$, ${\eqa}f$, and ${\lel}f$, ${\gel} f$, ${\sim}f$
denote ${<}f{+}O(1)$, ${>}f{-}O(1)$, ${=}f{\pm}O(1)$, and ${<}f{+}O(\|f
{+}1\|)$, ${>}f{-}O(\|f{+}1\|)$, ${=}f{\pm}O(\|f{+}1\|)$, respectively.
 \\
 $Q(G)$ is the probability of a set $G$ or \trm {mean} $\sum_xQ(\{x\})G(x)$
of a function $G$ by a distribution $Q$.

We use a \trm{prefix} algorithm $U$: $U(p){=}x$ iff $U(p0){=}U(p1){=}x$.
Auxiliary inputs $y$ in $U_y$ are not so restricted.\footnote
 {All results below remain valid, of course, if
 relativized by giving $U$ an extra auxiliary input.}
$p$ is \trm {total} if $U$ halts on all $k$-bit $ps$ for some $k$. Our $U$ is
\trm{universal}, {i.e.} minimizes (up to $\eqa$) complexities $\K$, $\|\M\|$
below, and \trm {left-total} : if $U(p1s)$ halts, $p0$ is total.\footnote
 {$U'$ is turned into left-total $U$ by enumerating $p$ in order of
  convergence of $U'(p)$ and assigning them consecutive intervals
  $i_p\{0,1\}^\N$, $\|i_p\|=\|p\|{+}1$ shared by $p,q$ with $\|p\|{=}
  \|q\|,U'(p){=}U'(q)$; then $U(p'){\edf}U'(p)$ if $p'\in i_p\{0,1\}^*$.}
 $\ch(i){\edf}[U(i)\mbox{ halts}]$.

\trm {Complexity} $\K(x|y)$ is $\min_p\{\|p\|:U_y(p){=}x\}$.
$\M_v(G)=\sum_p2^{{-}\|p\|}[U(vp^-) {\ne}U(vp){\in}G]$ is \trm
{universal probability}. We omit empty $y,v$. $\|\M(\{x\})\|{\eqa}\K(x)$.
 \\
 $\I(x{:}\,y)\edf\K(x){+}\K(y){-}\K(x,y) \eqa\K(x){-}\K(x|(y,\K(y)))$
is \trm {information}. $\I(x{:}\,\ch)\edf\K(x){-}\K(x|\ch)$.

\trm {Rarity} (non-randomness) $\d(x|Q,v)$ is $\floor{\,|\log Q(\{x\})|\,}{-}
\K(x|v)$. $t_{Q,v}(x)=2^{d(x|Q,v)}$ is a rarity \mbox{\trm{$Q$-test}} i.e.,
$Q(t_{Q,v}){\le}1$ for any $Q,v$. It is the largest test, i.e., $t'=O(t)$ for
any lower-enumerable $t'_{Q,v}(x)$ $Q$-test for computable $Q_v(\{x\})$. $\ov v$
is a program for $v=U(\ov v)$; $\|\ov v\|\lea\l(v){\edf}\|v\|+\K(\|v\|)$.

\newpage
 \section {The Results.}

For $f(n){\in}O(n)$, $Q_v(\{x\}){=}U_x(v)$, we use a slice $\Ks_f(a)\edf\min_v
\{\|v\|{+}f(\d(a|Q_v,v))\}$ of Kolmogorov structure function, requiring
$Q_v(\St){=}1$ unlike \cite{Shen83}. $\Ks\edf\Ks_\l$. Low-$\Ks$ ({i.e.} random
under simple distributions) $a$, Kolmogorov called \trm {stochastic}. The other
$a$ are ``exotic,'' {i.e.} have high $\I(a:\ch)$:

\begin{prp}\label{P1} $\I(a:\ch)\gel\Ks_f(a)$.\end{prp}

\prf Let $U(vw){=}a$, $\|vw\|{=}\K(a)$, $v$ be total, $v^-$ be not.
Then $\|v\|{+}\|w\|=\K(a)\lea$ $\K(a|v)+\K(\|v\|){+}\|v\|$, so
$\|w\|{-}\K(a|v)\lea\K(\|v\|)$. Using $\M_v$ for $Q_{\ov v}$, gives
$\Ks_f(a){\lea}\, \l(v)+O(\|w\|{-}\K(a|v))\lea\l(v){+}O(\K(\|v\|))$.
Now, $\K(a|\ch){\lea}\K(\|v\|){+}\|w\|$, so
$\I(a:\ch)\gea\|v\|{-}\K(\|v\|)\gel\Ks_f(a)$.~\qed

Then we prove that all stochastic sets have simple (high $\M$) members:

\begin{lem}\label{L1} $\|\max_{x{\in}D}\M(\{x\})\|\lea
\l(\M(D))+\|\K(\|\M(D)\|)\|+\Ks(D)$.\end{lem}

\paragraph {Informal proof outline:} We enumerate a small (thus of low members
complexity) $L$, and a test $\tb(X)$, high for $X\subset\St{\setminus}L$
with $\M(X)\ge\M(D)$. This assures $\d(X|Q,v)>\d(D|Q,v)$, so $X\ne D$.
 \\
 We break inputs of $U$ into ${\approx}\M(D)/\d(D|Q,v) $-wide intervals $p\St$.
 \\
 In each interval with total $p$ we select one output $L_p{=}U(pp')$
 and update a $Q$-test $\tb_p(X)$.
 \\
 Its $\ln(\tb_p(X))$ accumulates $\M_p(X)$, until $\{L_r|r{\le}p\}$
 intersects $X$, upon which $\tb_p(X)$ drops to $0$.
 \\
 The test $\tb(X)=0$ if $\max_p\tb_p(X) <J\sim e^{\d(D|Q,v)}$, else
$\tb(X){=}J$. $L_p$ is selected to keep $Q(\tb_p)\le1$. This is possible
since the mean choice of $L_p$ does not increase $Q(\tb_p)$, and the
 \\
 {\bf minimal increase cannot exceed the mean}:
this is the {\bf key} point of the proof.

\paragraph {Formal proof:}Let $v,Q{=}Q_v$ minimize $\Ks(D)$, $i\edf\|\M(D)\|$,
$d\edf\d(D|Q,(v,i))$, $j{\eqa}\|d\|$, $J{\edf}e^{2^j{-}1}$. For all total
$p{\in}\{0{,}1\}^{i{+}j}$, we build inductively a list
$L{=}\{L_p{\in}U(p\St)\}$ and $Q$-tests $\tb_p$ $({=}\tb^L_p(X))$, using $L_p$
and $\tb'_p{=}\tb_{p{-}1}$ (or ${=}1$ if $p{=}0^{i{+}j}$): $\tb_p\edf\tb'_p$ if
$\tb'_p{\in}\{0,J\}$, else $\tb_p\edf\min\{J[L_p{\notin}X],\,e^{\M_p(X)}\tb'_p\}$;
$\tb\edf J[J{=}\max_p\tb_p]$. $L,\tb$ will be enumerable from $v,i,j$.
Let $L_{p,s}$ be $\{L_r|r{<}p\}$ with added $L_p{=}s$.

By $(1{-}a)e^a{\le}1$ for $a{=}\M_p(X)[\tb'_p(X){<}J]$, we get $\sum_s
\M_p(\{s\})\tb_p^{L_{p,s}}(X)\le(1{-}a)e^a\tb_{p{-}1}^L(X)\le\tb_{p{-}1}^L(X)$.
So the mean $\sum_s\M_p(\{s\})Q(\tb_p^{L_{p,s}}) {\le}\, Q(\tb_{p{-}1}^L)$;
thus $Q(\tb_p^{L_{p,s}}){\le}\,Q(\tb_{p{-}1}^L)$ for some $s{\in}U(p\St)$.
Such choices of $L_p{=}s$ assure $Q(\tb_p^L){\le}1$ for all total $p{\in}\{0,1\}
^{i{+}j}$, so $\tb_p^L$, $\tb$ are $Q$-tests.

$\sum_p\M_p(D)=2^{i+j}\M(D)>2^j{-}1$, so $\tb(D){=}J$ if $D\subset\St
{\setminus}L$. Then $D$ intersects $L$, as otherwise $\|\tb(D)\|=\|J\|>1.44
(2^j)$ and $d>\d(D|Q,(v,i,j))-\K(j){-}O(1)>\|\tb(D)\|{-}\K(j){-}O(1)>d$.

So, for $s{\in}L$, $\K(s){\lea}\,i{+}j{+}\K(i,j,v)\lea i{+}\K(i){+}\|\K(i)\|
{+}\Ks(D)$, as $j{\eqa}\|\d(D|Q,v)\|$ or $j{\lea}\|\K(i)\|$.~\qed

\begin{thr} $\min_{x{\in}D}\K(x){\eqa}\|\max_{x{\in}D}\M(\{x\})\|\lel
\|\M(D)\|{+}\I(D:\ch)\sim\!\min_{x{\in}D}\I(D:x){+}\I(D{:}\ch)$.\end{thr}

\prf $\I(D:x){\eqa}\K(x){-}\K(x|(D,\K(D))\gel[x{\in}D]\,\|\M(D)\|{=}i$. The
latter is achieved by a distribution $\mu_{(i,D)}(\{x\})=\M(\{x\})2^i[x{\in}D]$.
So, the Lemma and Proposition~\ref{P1} complete the proof.~\qed

\paragraph {Acknowledgments.}
 {{Besides Samuel Epstein,}}
 much gratitude is due to Margrit Betke, Steve Homer,
 Paul Vit\'anyi, and Sasha Shen for insightful discussions.

 \end{document}